\documentclass[conference]{IEEEtran}
\ifCLASSINFOpdf
\else
\fi

\PassOptionsToPackage{bookmarks={false}}{hyperref}

\usepackage{amsmath}
\usepackage{booktabs}
\usepackage{caption} 
\usepackage{graphicx}
\usepackage[draft]{hyperref}

\usepackage{capt-of}
\renewcommand\IEEEkeywordsname{Keywords}

\begin{document}
%
\title{Understanding Dhaka City Traffic Intensity and Traffic Expansion Using Gravity Model}

\author{\IEEEauthorblockN{Md Abu Sayed\IEEEauthorrefmark{1}, Md Maksudur Rahman\IEEEauthorrefmark{2},  Moinul Islam Zaber\IEEEauthorrefmark{3}, Amin Ahsan Ali\IEEEauthorrefmark{4}}
\IEEEauthorblockA{Department of Computer Science and Engineering ,University of Dhaka. \\
Dhaka, Bangladesh. \\
Email: \{\IEEEauthorrefmark {1}2012-81-2031, \IEEEauthorrefmark{2}2012-71-2014\}@students.cse.univdhaka.edu,
\{\IEEEauthorrefmark{3}zaber, 
\IEEEauthorrefmark{4}aminali\}@du.ac.bd }}

\maketitle

\begin{abstract}
Analysis of traffic pattern recognition and traffic
congestion expansion in real time are one of the exciting and
challenging tasks which help the government to build a robust
and sustainable traffic management system specially in a densely populated city like Dhaka. In this paper, we analyze the traffic intensity for small areas which are also known as junction points or corridors. We describe Dhaka city traffic expansion from a congestion point by using gravity model. However, we process
real-time traffic data of Dhaka city rather than depend on survey and interview. We exactly show that traffic expansion of Dhaka city exactly follows gravity model. Expansion of traffic from a congestion point spreads out rapidly to its neighbor and impact of congested point decreases as the distance increases from that congested point. This analysis will help the government making a planned urbanized Dhaka city in order to reduce traffic jam.
\end{abstract}



%
\IEEEpeerreviewmaketitle

\section{Introduction}

Traffic pattern recognition and formulation using real time
road traffic data are one of the most essential and challenging tasks considerably related and affecting to our day-to-day life. A better traffic pattern recognition system helps individual by saving their valuable time to reach the destination in time. It also helps government to minimize human sufferings from acute traffic jam problem through analyzing traffic pattern rather than depend on trivial traditional traffic management system which does not take adequate measure to give remedy from acute traffic jam problem instead increases traffic jam because of limited knowledge of the whole picture of traffic
jam, traffic patterns, and traffic expansion. A better traffic pattern recognition system can overcome the shortcomings of the trivial whistle based traffic management system. Most important thing for the better traffic recognition system is to understand traffic situation, to understand traffic behavior of the area from real-time traffic data rather than survey data.

Dhaka is one of the fastest-growing city in the world. Its
population was 3 million in 1971 and today it is 18 million.
This unprecedented urban growth has made Dhaka city more
densely inhabited than any other major city. Dhaka now one
of the of the worst traffic congested cities in the world.
Two major factors contribute to Dhaka city current traffic
congestion: lack of planning and preparation over previous
decades, and an over-reliance on cars due to a less amount of public transportation system. Even though there are 33 times more cars than buses in the city, around 13 percent passengers use cars, while 49 percent passengers use buses. Today, the average traffic speed in Dhaka city is 6.4 kph, but it may fall to 4.7 kph by 2035 if vehicle growth continues at its current rate. People spend an average two and half hours in traffic activities daily of which one and half hours are eaten up due to traffic congestion \cite{star}.

Bangladesh government have taken some initiatives to control traffic congestion in Dhaka city. For traffic signaling, the government is planning to introduce remote controlled traffic signal system in Dhaka city instead of following traffic polices hand language, or trivial whistle based traffic control system.In 2014-15, the government took a Revised Strategic Transport Plan (RSTP), which proposed building five metro rail lines, two rapid bus routes, and 1,200 kilometers of new roadways double the existing major road network including six elevated expressways and three ring roads. RAJUK has undertaken some steps for mitigating the suffering of traffic jam such as constructing the extension of some busy roads, widening of link roads. RAJUK has published the future planning of transportation system dividing the whole area into six zones as it’s Dhaka Structure Plan (2016-2035) \cite{rajuk}. In order to minimize the traffic intensity, Bangladesh Bridge Authority has intended to deliver a project for the construction of approximately 23 km of Elevated Expressway in the northern part of Dhaka city. Japan International Cooperation Agency (JICA) has conducted a preparatory survey called Dhaka Urban Transport Network  Development Study (DHUTS) in Bangladesh. They integrate urban development plan of Dhaka Metropolitan Area with the formulation of the Urban Transport Network Development Plan \cite{7}.

To find the reasons behind traffic congestion, we need
to understand traffic condition and traffic expansion. Traffic expansion information helps us to build a structured Dhaka city, especially where to construct a flyover or where to give a special concern. But, all recent studies of traffic in Dhaka city do not give much focus on this aspects. They mainly depend on interview and field surveys that do not give you a complete picture of traffic congestion. The main problem of survey data is that its become older and road condition also changes rapidly. For this reasons, we have focused on understanding traffic intensity and traffic expansion using the real-time traffic data set we have. Understanding traffic situation and traffic behaviors are also essential for analyzing road traffic, to predict vehicle expected speed for analyzing causing of road accidents, to predict road traffic intensity according to the number of vehicles on the road and number of vehicles crossing the road in a unit time. By monitoring traffic condition in real time we can suggest alternative options instantly to a passenger, to a bus driver, to a traffic police and also helps network analyst to plan a structured traffic free or traffic less city.

In this paper, to understand Dhaka city traffic condition we
plot some graphs and traffic expansion by using the gravity
model. We show that road traffic expansion of any city exactly follows the Newtons universal law of gravitation which also known as gravity model. The Gravity Model is a model used to estimate the amount of interaction between two cities. For estimating interactions between two cities it needs populations of the city, the distance between two cities which are used as parameter or input for the gravity model. By calculating impact gravity model can tell migration rate from one city to another city, commodity flow from one city to another city. Gravity model inherently told us that if the population of two
cities increases or distance between two cities decreases. If the population of two cities is small or distance between two cities is large then the amount of interaction between two cities is definitely low. There are many variations of the gravity model. The gravity model is commonly used in analyzing traffic and economic activities\cite{5}. We had to tackle some challenges.
Firstly, processing a huge amount of real-time road traffic
data of a city for extracting information to visualize traffic on every road segment or visualize traffic on a small area of the city is always challenging. Extracting precise information from this huge amount of data is also challenging. Secondly, at the time of processing data, we have tried to found some important observations from data such as variation of traffic of a road segment on weekend and weekdays, variation of traffic on working hour than any other hour. Thirdly, for simulating gravity model we need to have precise information and we also find out carefully how roads information are organized in our data set. Another most challenging task is to defining the gravity model parameters. Because gravity model takes a huge variety of parameters according to on what data we run gravity model, what types of things we want to simulate
by using gravity model. For this, we need to answer some
questions. What is the traffic intensity of a road segment? What is the traffic intensity of the junction point?What variation of gravity model should we use? Why or why not gravity model parameters value remain between some specified range?

In this paper, we use real-time traffic data of Dhaka city
for analyzing its traffic behavior, for analyzing its traffic expansion.There is no past research which analyzes Dhaka city traffic intensity and traffic expansion from real-time traffic data, while they depend on survey data. We collected Dhaka city traffic data from authorized repository. Before starting analysis we observe the whole data set carefully to figure out what type of information we exactly have in our data set regarding traffic of Dhaka city and how they organized. We also carefully examine features from our data set so that we got precise information about Dhaka city traffic from our data
set. We mainly figure out all road connectivity points with
each other in Dhaka city by plotting their GPS coordinates on Google map\footnote{\url{https://www.google.com/maps/}}. We select the corridor and junction point based on the basic design study report on Dhaka Urban Transport Network Development Study (DHUTS) conducted by Japan International Cooperation Agency (JICA). On the junction point and corridor, we try to understand Dhaka city traffic intensity and traffic expansion from a junction point or from a corridor.

\section{Related Work}

Many analyses have been done regarding traffic pattern modeling and the usage of gravity model. Gravity model illustrates the socio-economic relations using traffic behavior between different points or junctions. However, traffic pattern modeling helps us to understand the factors that change traffic behavior.

In \cite{1} the authors explain that management of road traffic needs high-tech computerized solutions rather than manual methods that such as traffic policemen, traffic lights, and safety cameras, whistle based traffic system. Collection and analysis of road traffic data is a major requirement for establishing of traffic conditions on any given road segments. This paper focuses on the use of the Global Positioning System (GPS) technology for traffic data collection and analysis of traffic conditions. In this research project, researchers have developed a GPS data receiver application and traffic analysis system that collects GPS traffic data and provide the ability for monitoring and analyzing traffic scenarios on the roads, for instance, the speed of traffic. The system also provides planners on the road usage patterns for decision making.

In \cite{2} gravity model is used for analyzing the distributions of urban truck trips and urban commodity flow. Firstly they discussed the applicability of gravity model on these applications and secondly they discussed differences between track trips and commodity  flow. Their result shows that gravity model is applicable for analyzing the distribution of urban truck trips and also applicable for analyzing those commodities which source and destination never limited to a small number of locations.

In \cite{3} the authors investigated traffic flows on Korean
Highway system which includes both public transportation and
private transportation information. They showed that traffic
flow $T_{ij}$ between city $I$ and $J$ follow the Newtons universal law of gravitation. $T_{ij}$ follows $P_{i}\times  P_{j}$ / $r_{ij}^2$ where $P_{i}$ and $P_{j}$ are the
populations of the city $I$ and $J$ and $r_{ij}$ is the distance between city $I$ and $J$. They also showed that Korean highway system both public and private transportation system has heavy tail even tough they are uniform and homogeneous than a random network.

In \cite{4} the authors proposed a spatial network model for transportation system by considering optimal expected traffic. Optimal expected traffic is the prediction of traffic flow between two nodes is calculated by gravity model equation. They proposed an improved formula for calculating expected traffic which is $W_{ij}$ = $K\times M_{i}^a\times M_{j}^a$ /  $D_{ij}^b$, where parameter a, b is used to control the fitness and geographical constraints. They also showed the relationship between expected traffic and real traffic.

In \cite{5} the authors showed the effect of variation of a gravity model basic parameter such as distance, travel time, commodity  flow exponent. They also showed that variation of the exponent balanced with the urban context. They relate variables with a huge number of exponents using data of 1967 commodity flow of United States. They used multiple regression models for the derived exponent. For the better fit of the exponent, they use root mean squares. Most implication factor of their research is that their exponent result is compatible with another scenario.

In \cite{6} the researchers tried to use mobility model to examine underlying stuff of human mobility. They also use gravity model for analyzing human mobility patterns on various transportation systems. They used traffic of various transportation systems which is a function of distance and populations as parameters or variables of the gravity model. They also focused on applicability and variability of gravity model on intra-urban mobility.

Dhaka Urban Transport Network  Development Study (DHUTS) in Bangladesh is a preparatory survey conducted by Japan International Cooperation Agency (JICA) \cite{7}. Their main objectives are to integrate urban development plan of Dhaka Metropolitan Area with the formulation of the Urban Transport Network Development Plan. According to their plan, they draw the priority basis general outline of the urban transport projects. On their study, they proposed an alternative transport network development for planned Dhaka city up to 2025. In their study, they propose adequate step for overall traffic scenario of a road segment in Dhaka city. They took three steps for various types of the road segment. Their steps are Do Nothing scenario, Do minimum scenario and Do maximum scenario. The DHUTS model reveals that under do-nothing condition only with the on-going projects. Do minimum scenario, Do maximum scenario give corridor information of Dhaka city on a priority basis. From Do minimum scenario, Do maximum scenario we collect corridor information on which corridor we actually run gravity model.

\section{Data Description}

In this section, we describe our data set and the types of it. We obtained the data from a private company named Gobd\footnote{\url{gobd.co} analyzes traffic pattern and suggest routes of Dhaka city}. The traffic intensity is calculated using the GPS information of taxicabs that run through the Dhaka city. The characteristics and the factors (i.e., speed, altitude, road width proxy) are obtained by using Open Street Map.

We have total 15 days data from 1-9-2015 to 15-9-2015. In
our data set, we have total 16,23,280 rows and 18 columns
or 18 features. Every entry has a unique object identifier.
Every road has a human readable name which column name is OsmName. The Same name is used for forwarding and
backward direction of the road. Every road segment has a list of GPS coordinates including starting and ending point. Every road segment has unique open street map identifier. Every road segment has also unique open street source identifier and unique open street target identifier. Every road segment has also its created time which data type is date time. Every road segment has also classified into a class based on business or expected traffic such as highway traffic or street traffic. Every road segment has also its overall length. We also have vehicles speed on every road segment. We also have overall cost information of every road segment. In our data set, many road segments have the same OsmName. Traffic density of a road segment in our data set represented by intensity which takes real numbers value from 0 to 1. In our data set intensity
of a road is calculated from longitude and latitude information of that road, vehicles speed on the road and class of that road. Some of the columns are given below:

InferredIntensity: Inferred Intensity is a continuous variable which takes real numbers. The intensity of
traffic jam on the road segment within 0.0 to 1.0 range. But the Inferred Intensity is not an average of the forward and backward traffic of a road. It is either forward traffic intensity information or backward traffic intensity information. Inferred Intensity value close to
1.0 indicates the high traffic intensity on those roads. Inferred Intensity value close to 0.0 indicates the low traffic intensity on those roads. In our data set the intensity of a road is calculated from longitude and latitude information of that road, vehicles speed on the road and class of that road\footnote{\url{https:/wiki.openstreetmap.org/wiki/Using_OpenStreetMap/}} [Data Type: Float].

TimeStamp: Time stamp, when the record was created. A time stamp is a Date Time object. From Time Stamp, we can extract a day, hour, minute, second [Data Type: Date Time].

WayID: Unique identifier of the road segment. For every road segment, its unique. Every road forward and backward directions are classified into several segments and for every segment, we have a unique Way ID [Data Type: Integer].

OsmID: Open Street Map identifier for the road segment, for visualizing in Java Open Street Map (JOSM). Every road is classified into several segments and for every segment, our data sets have a unique OsmID. It is also true for bidirectional roads which every direction are classified into several segments and for each segment, our data sets have unique OsmID [Data Type: Integer].

OsmSource: OsmSource is the Starting point of the road segment in Open Street Map. It is also identifiers of the starting point of every road segments [Data Type: Integer].

OsmTarget: OsmTarget is Ending point of the road segment in Open Street Map. It is also identifiers of the ending point of every road segments [Data Type: Integer].

Class: Class of the road segment based on business/expected traffic flow (e.g. Highway vs segment street) [Data Type: Integer].

We have plotted all roads GPS coordinates on Google map to find connectivity information among roads segment that is needed for examining the effect of congested road segment
using gravity model. After plotting GPS coordinates on Google map we find that many of the road segments have several OSM name and we consider all of this several OSM name as the same road segment. This information helps us to understand Dhaka city traffic.

\section{Methods}
In this section, we describes our analyses and the procedures we follow. How, we use the gravity model using different junctions and corridors are also mentioned here.

\subsection{Selecting Junction Point and Corridor}
\begin{figure}[!h]
  \centering
  \includegraphics[scale=0.5]{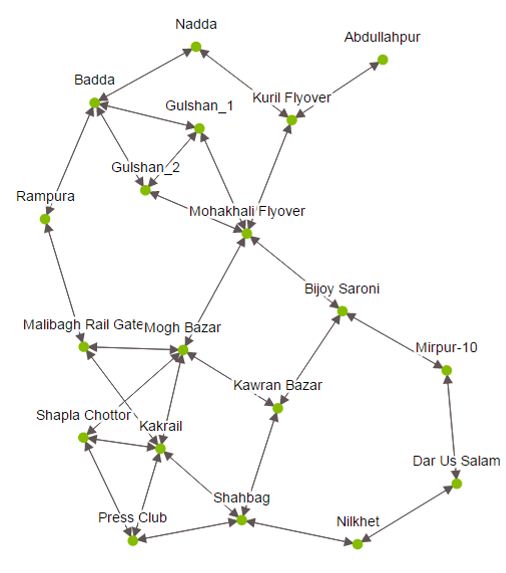}
  \caption{ Connectivity map of all junction point. After plotting all GPS point on google map we found this connectivity on collected junction point from our data set.Here all roads are bidirectional to reach from one junction point to another.}
  \label{fig:dendo1}
\end{figure}

In this paper, we explore traffic pattern, traffic intensity and traffic expansion from a junction point or on a corridor. A junction point is a collection of all incoming and outgoing road segment on that point. A corridor is defined only by a backward or forward direction traffic along through that corridor. In junction point, we do not consider forward and backward direction traffic separately, but in the corridor, we consider it separately. We collect junction point and corridor information from Dhaka city urban transport development study of JICA \cite{7}. We have 19 junction points. Figure \ref{fig:dendo1} shows the image of our all junction point.

\begin{figure}[!h]
  \centering
  \includegraphics[scale=0.4]{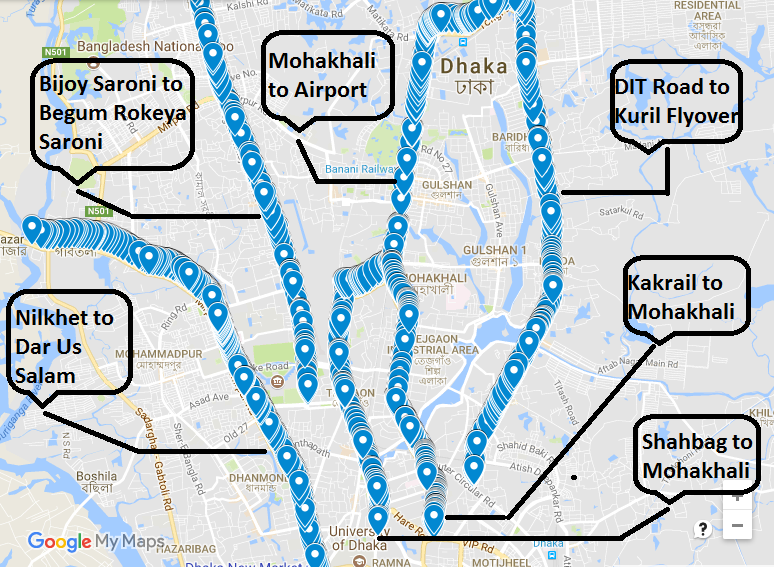}
  \caption{All Corridor information on Google map. 
  }
  \label{fig:dendo2}
\end{figure}

We have 6 corridors named as Shahbag to Mohakhali, Nilkhet to Dar Us Salam, Kakrail to Mohakhali, Bijoy Saroni to Begum Rokeya Saroni, Mohakhali to Airport, DIT to Kuril Flyover. Figure \ref{fig:dendo2} is the image of our corridor.
\\

\subsection{Understanding Dhaka City Traffic}
In our data set, all road segments do not have enough amounts of data, but some have adequate data. First, we calculate all road segment average traffic information and all junction average traffic from whole data which cannot give us a complete picture about road segment and junction point traffic. Average hour wise road segment and junction point traffic information from whole data gives better information than average road segment and junction point traffic information. But every days hour wise traffic information of a road segment or junction point gives more precise information than the average hour wise traffic information of road segment or junction point from the whole data. It is true when you have a huge amount of data. 

But if we were seeing every day half hour wise traffic intensity of road segment or junction point, which might give us more precise information. But in our data set, we do not have every moment data of all road segment or junction point. But sometimes this types time interval may introduce overfitting problem. Also most of the times, traffic created and expansion needs more time. If data sets have several month data then week wise hour traffic data or week wise every day traffic helps us to visualize more accurately rather than depend on some days data. To visualize and understand traffic intensity we plot some day wise, hour wise and some aggregated graph of junction and corridor.

\subsection{Analyzing Traffic Expansion using Gravity Model}
For exploring traffic expansion from a junction point or on a corridor we have used gravity model. We use the gravity model to analysis impact of one junction point to another junction point. We also measure the impact on a corridor in both forward and backward direction. Here we exactly prove that impact of one junction on another junction is increased if the intensity of two junctions are increased or distance between two junctions are decreased and decreased if the intensity of two junctions are decreased or distance between two junctions are increased. Here is our law of gravity model.
\[ IMPACT_{ij}
= G \frac{I_{i} I_{j}}{d_{ij}^2}  \tag{1} \label{eq:1}  \]
Here, $IMPACT_{ij}$ indicate the impact of junction i to junction j on a specific hour. $I_{i}$ is average intensity of junction i on a specific hour t. $I_{j}$ is average intensity of junction j on t+1 hour. $d_{ij}$ is distance between two junction i and j. For the value of G, we use 1.
\section{Result}
The findings of our analyses are described is this section. We compare different situations (i.e., overall, days of week) for understanding traffic intensity pattern.

\subsection{Overview of Dhaka City Traffic}
We find that the traffic intensities varies with respect to time throughout a day. We examine hour wise traffic intensity variation. We take the average hour wise traffic intensity of all days. Then we plot the hour wise intensity graph which represents the traffic intensity pattern of whole Dhaka city in Figure 3. There the X axis represents the time slots (hour) and Y axis represents the corresponding traffic intensity in that time slot.

\begin{figure}[!h]
  \centering
  \includegraphics[scale=0.15]{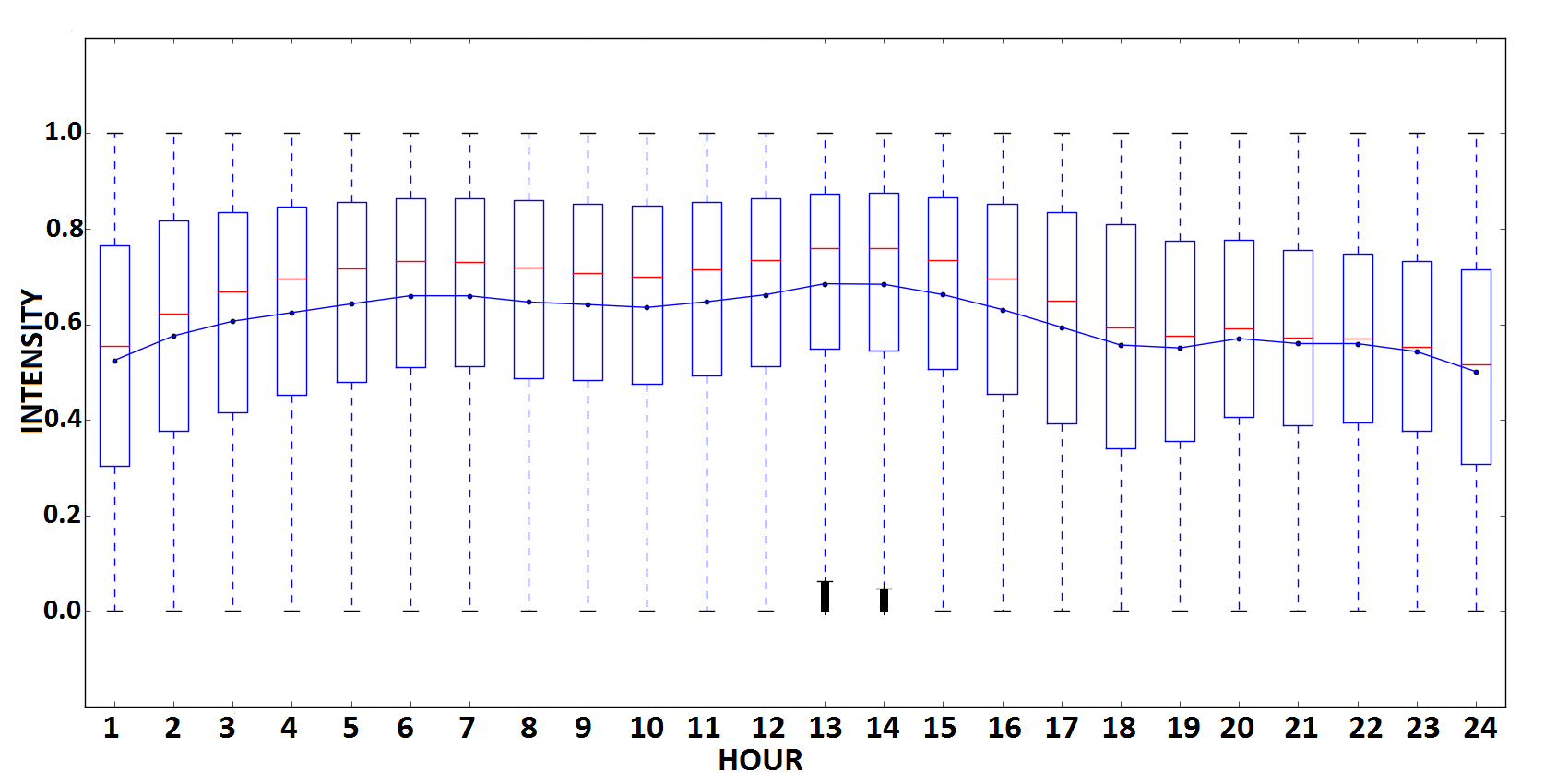}
  \caption{Dhaka city traffic intensity pattern over day. Traffic intensity is relatively high in working hours }
  \label{fig:dendo}
\end{figure}

From Figure 3, we found that Dhaka suffers from heavy traffic intensity in most of the hour of a day, specially on working hour. At night from 10.00 pm to 05.00 am Dhaka city traffic intensity is relatively low.

\begin{figure}[!h]
  \centering
  \includegraphics[scale=0.3]{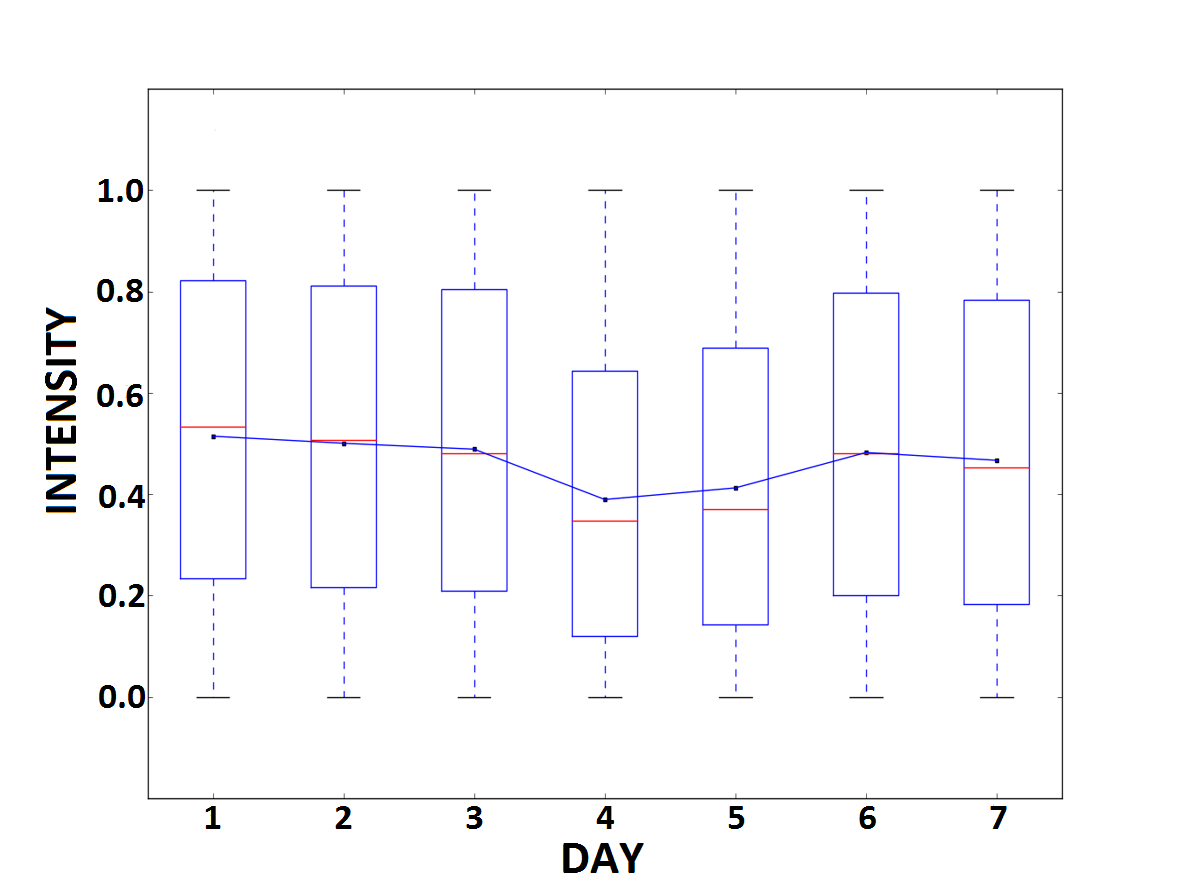}
  \caption{Day wise intensity graph.}
  \label{fig:dendo}
\end{figure}

We also plot some day wise intensity graphs. From Figure
4, we see that weekend has less traffic than another weekday. Here, day 3 represents Friday and day 4 represents the traffic information of Saturday. It is quite clear from the figure that traffic intensity pattern is quite low on weekends (Friday and Saturday) compared with other working days. 

From Figure 5 (hour wise graph aggregated over two
weeks), we see that the two weeks have same traffic intensity in most of the hours. This graph also shows variability between two weeks. Too much variability of hour wise intensity graph also be needed for further analysis.

\begin{figure}[!h]
  \centering
  \includegraphics[scale=0.27]{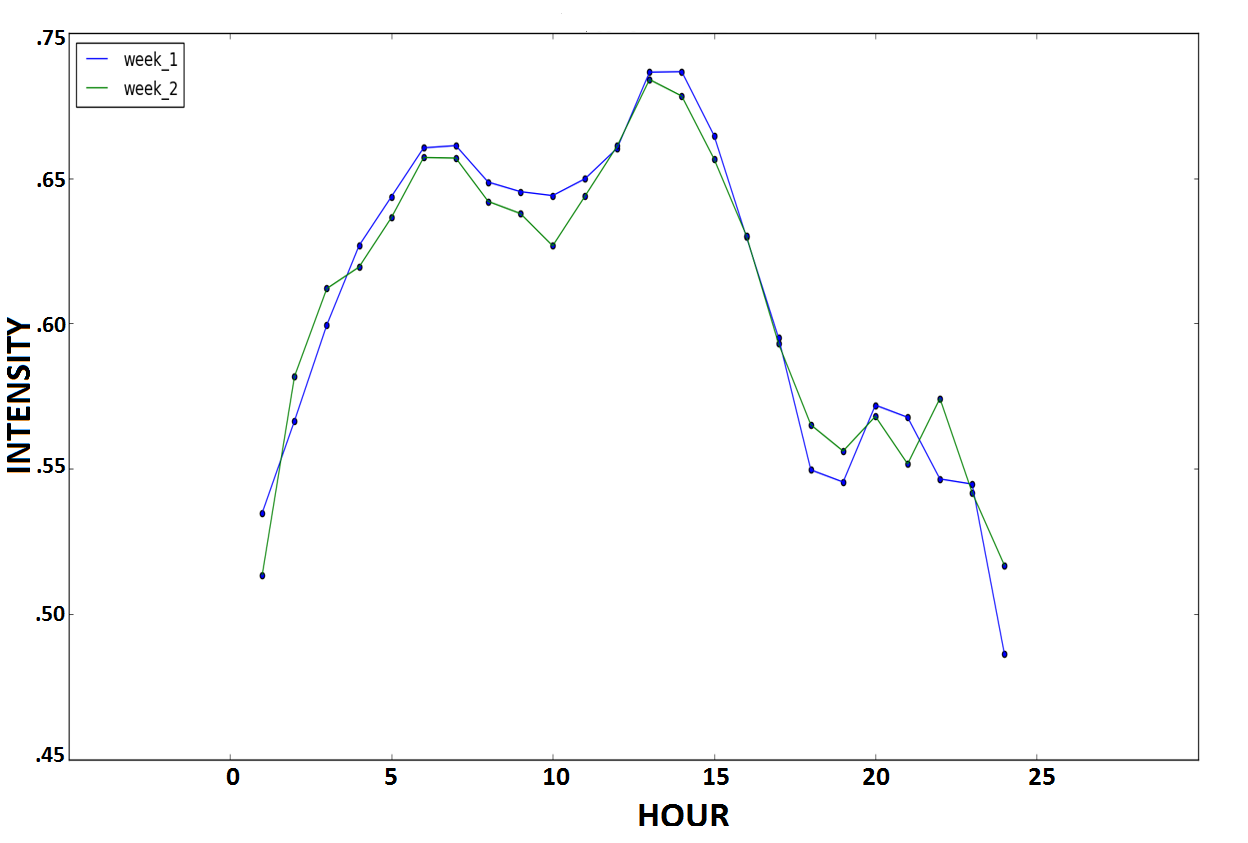}
  \caption{Aggregate of two week data for whole Dhaka City.}
  \label{fig:dendo}
\end{figure}

\subsection{Analyzing Traffic Expansion using Gravity Model on a Junction Point}

In our work, we define Intensity of a junction point by
taking average of the total amount of incoming and outgoing
road segment traffic on that junction. For finding precise information, we also calculate all day hour wise average intensity and every day hour wise average intensity of every junction. We also define hop by the number of road segment that we need to reach from source to destination.

\begin{figure}[!h]
  \centering
  \includegraphics[scale=0.218]{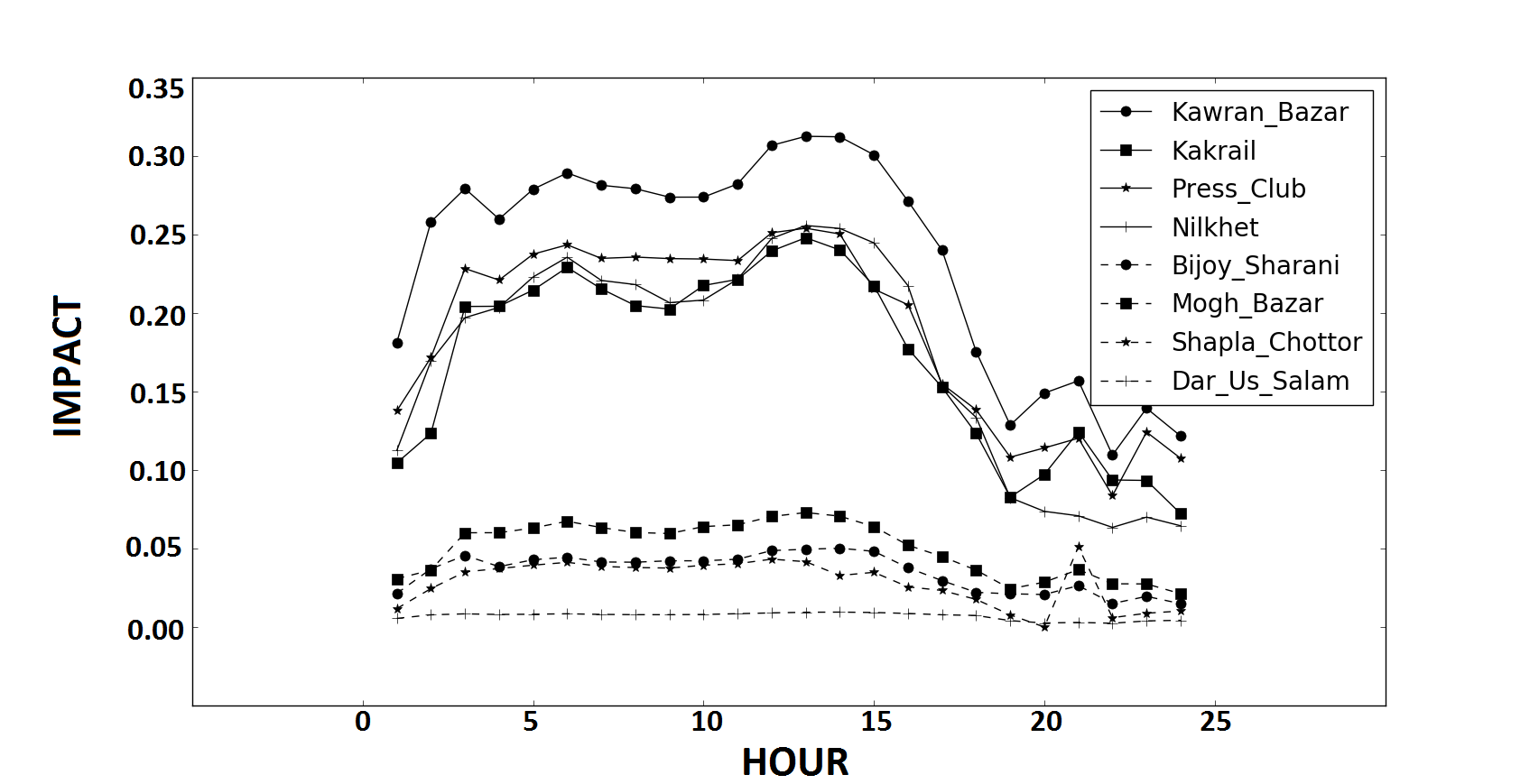}
  \caption{Impact of Shahbag on its neighbour over hop1(continuous line) and hop2(discontinuous line).}
  \label{fig:dendo}
\end{figure}

We calculate the impact of one junction on another junction
on a particular hour by only seeing Intensity of that junction on an observing hour and the intensity of impacted junction on the next hour. Because if one junction point becomes congested, all other connected junction will be subsequently congested within one hour or less time. But we only see what impact on its neighbor link after one hour later.

From Figure 6 we see that Shahbagh pushes more impact on
hop1 than hop2 because of their high distance from Shahbagh
and low intensity at that time and its also push more impact
on all of its neighbor junction points at afternoon rather than any other time and also it pushes less impact on all of its neighbors at 11 pm to 6 am.

\subsection{Analyzing Traffic Expansion using Gravity Model on a Corridor}
We examine the traffic expansion on a corridor using gravity model. We choose Shahbagh to Mohakhali corridor and examined the impact of gravity model through backward and forward direction.

\begin{figure}[!h]
  \centering
  \includegraphics[scale=0.218]{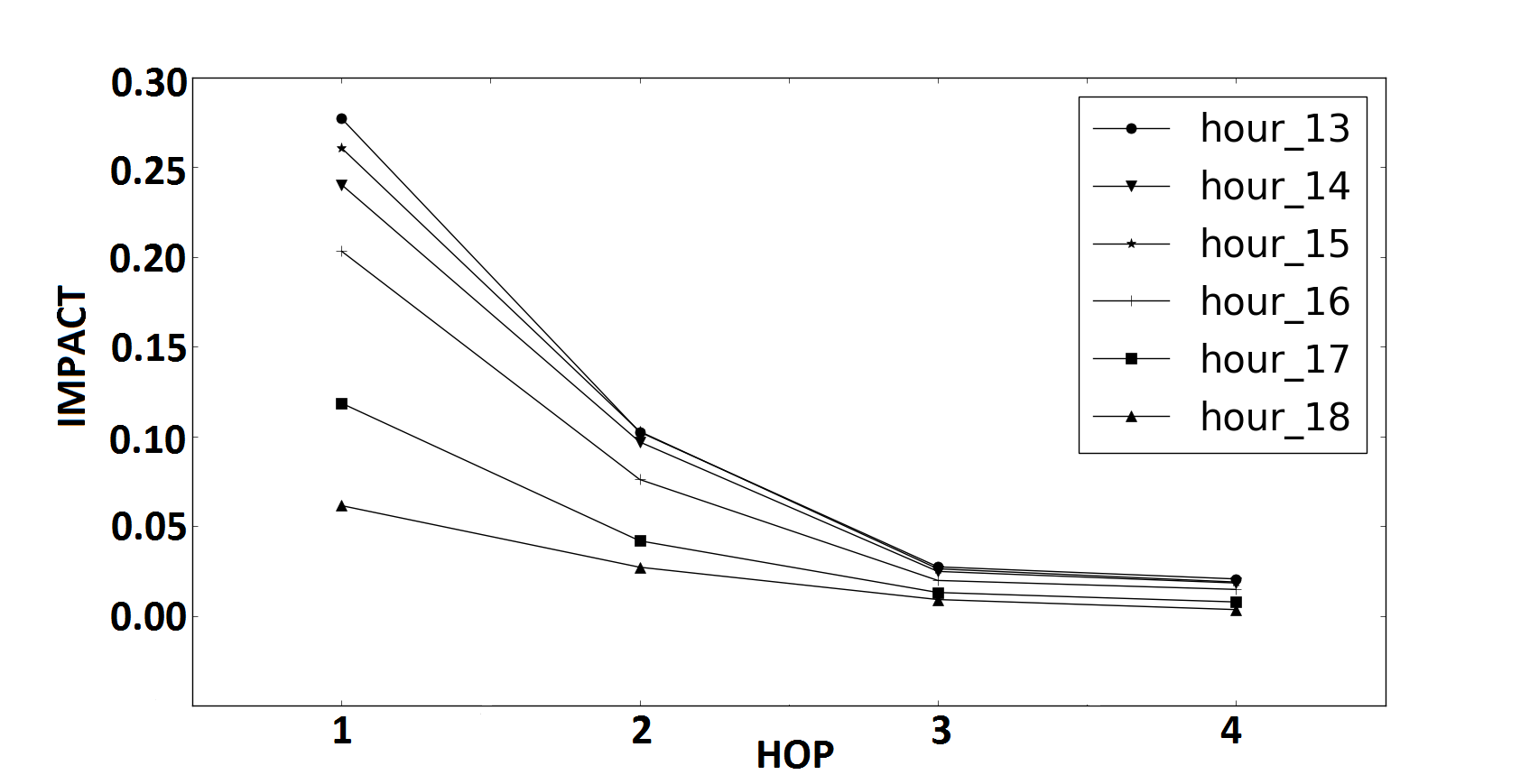}
  \caption{ Impact of road segment on a corridor .}
  \label{fig:dendo}
\end{figure}

Here above Figure 7 shows the hour wise impact of Shahbagh on its neighbor along Shahbagh to Mohakhali corridor. From this figure, we see that Shahbagh  pushes more impact on Kawran Bazar(hop1) and less impact on Mohakhali(hop4) because of their far distance and less intensity. It is also true for hour wise impact of Mohakhali on its neighbor along Mohakhali to Shahbagh corridor. Here, We show traffic expansion on Shahbagh to Mohakhali corridor. It is also true for any other corridor. But, sometimes on some corridor, some junction point push much impact on hop2 than hop1 because of hop2 less distance than hop1.

\section{Conclusion}
In this paper, we process a huge amount of traffic data for extracting information which needs for understanding Dhaka city traffic intensity and understanding Dhaka city traffic expansion. We give a very precise idea how traffic expansion from a congested point in Dhaka city and on what factor its actually depend. The main impact of our work is that it gives precise information about Dhaka city traffic and also gives a clear idea about how traffic spread out from a junction point or on a corridor. This information is helpful for government to build better traffic management system and also help urban development authority to build a structured city.
 
Traffic intensity information of a junction point or a corridor gives a complete picture to travelers and traffic network analyst. This information helps travelers to select a route to reach their destination, to estimate travel time to reach their destination. It also helps to select an alternative route for arriving his destination or select any other option. A network analyst can propose a better traffic management system by using this information. A network analyst can find out all shortcoming of existing traffic management systems and also they can measure travelers sufferings in the existing traffic management system.

In our work, the main hurdles were data availability of some corridors, the outlier on some hour wise intensity which is created by data availability problem, missing value of human readable open street map name.

However, we use the simplest version of the gravity model. In the future, we may extend our work by using another version of gravity model simplifying our gravity model parameter. We also mention the data availability problem of a corridor. In future, if we have a huge amount of data of all corridor then we will give a more exact picture of Dhaka city traffic intensity and traffic expansion.






%

\end{document}